# QOS EVALUATION OF HETEROGENEOUS NETWORKS: APPLICATION-BASED APPROACH

Farnaz Farid, Seyed Shahrestani and Chun Ruan

School of Computing, Engineering and Mathematics, Western Sydney University, Sydney, Australia
farnaz.farid@westernsydney.edu.au, s.shahrestani@westernsydney.edu.au, c.ruan@westernsydney.edu.au

### ABSTRACT

*In this paper, an application-based QoS evaluation approach for heterogeneous networks is proposed. It is possible to expand the network capacity and coverage in a dynamic fashion by applying heterogeneous wireless network architecture. However, the Quality of Service (QoS) evaluation of this type of network architecture is very challenging due to the presence of different communication technologies. Different communication technologies have different characteristics and the applications that utilize them have unique QoS requirements. Although, the communication technologies have different performance measurement parameters, the applications using these radio access networks have the same QoS requirements. As a result, it would be easier to evaluate the QoS of the access networks and the overall network configuration based on the performance of applications running on them. Using such application-based QoS evaluation approach, the heterogeneous nature of the underlying networks and the diversity of their traffic can be adequately taken into account. Through simulation studies, we show that the application performance based assessment approach facilitates better QoS management and monitoring of heterogeneous network configurations.*

## 1. INTRODUCTION

The advancement and proliferation of modern wireless and cellular technologies have changed the way people work and communicate. By 2018, the data traffic over mobile networks is expected to reach 15.9 exabytes per month, with 69 percent of that consisting of video. There will be over 10 billion mobile-connected devices by 2018, which will exceed the world's expected population at that time [1]. To deal with this growing number of devices and the massive increases in traffic, the networks are moving towards an all-heterogeneous architecture. Any heterogeneous network constitutes of different communication technologies. These technologies have distinct bandwidths, coverage area, and operating frequencies. Their QoS characteristics, such as delay, throughput, and packet loss, as well as usage and implementation costs also differ from each other. As a result, the adaptation of heterogeneous network-based architecture for the provision of different applications especially multimedia applications faces significant challenges. Among these challenges, QoS-related issues such as the effective QoS evaluation, management, and monitoring still top the list [2].

Managing QoS for video or voice applications over heterogeneous networks is a challenging task. A research from Nemertes shows that the companies invest a significant amount of their budget to manage VoIP applications over these network architectures. For small enterprises, the annual costs range from $25,000, and for global enterprises this cost is around $2 million [3]. Therefore, the enterprises need to dedicate a lot of their effort to ensure service quality at every level of the network. System downtime is another challenge for businesses, which could often happen due to poor network management and monitoring. According to Gartner research, the hourly cost of system downtime for large enterprises was $42,000, with a typical business on average, experiencing 87 hours of downtime per year [4]. As a result, the QoS of any service-based network should be monitored, managed, and evaluated on an ongoing basis.

In this paper, we introduce the concept of unified metric measurement functions that can help with assessing the application-based performance of heterogeneous networks. By taking the relevant performance-related parameters into account, these functions quantify the underlying network and the application-related QoS with a numerical value. The proposed approach considers the effects of the QoS-related parameters, the available network-based applications, and the available Radio Access Networks (RANs) to characterize the network performance with a set of three integrated QoS metrics. The first metric denotes the performance of each possible application in the network. The second one is related to the performance of each of the radio access networks present in the network. The third one characterize the QoS level of the entire network configuration. The core of this method is considering the effects of different application and radio access networks on the QoS of heterogeneous networks.

The rest of the paper is organized as follows: Section 2 discusses the background and motivations of this work. Section 3 illustrates the concept of unified QoS metric. Section 4 presents the application weight calculations in detail. Some simulation studies and result analysis are illustrated in section 5. The impact and the significance of the applications for QoS analysis are then discussed in Section 6. The last section gives the conclusions and proposes the future works.

## 2. BACKGROUND AND MOTIVATIONS

QoS evaluation in heterogeneous networks has been an active area of research [5, 6]. Most of the existing research focuses on the partial QoS evaluation of a heterogeneous network by deriving the performance level of a single access network and a single application present within that environment. Also, different studies have come up with various performance metrics for QoS evaluation of these networks. The conventional methods do not consider the performance of all the applications running on a network. For example, if there are voice and video conferencing running over a UMTS network, these methods do not include the performance analysis of these applications to quantify the overall network QoS. Furthermore, there is no unified metric to quantify the QoS of a network, which considers the performance of all the access networks present in it. For example, in a heterogeneous environment, there are three access networks, such as UMTS, WiMAX, and LTE. At present, no unified metric can represent the performance of this network configuration using the QoS-related parameters of these access networks.

Multi-criteria decision-making (MCDM) or Multi-Attribute Decision Making (MADM) algorithms have been widely employed in the area of the heterogeneous networks from vertical handover perspectives [7, 8]. The most common criteria, which are considered during this ranking process, are service, network, and user related [9]. These can include received signal strength, type of the service, minimum bandwidth, delay, throughput, packet loss, bit error rate, cost, transmit power, traffic load, battery status of the mobile unit, and the user's preferences. To facilitate the combining of these attributes into a single value, based on their relative importance, a weight is assigned to each attribute.

The weights for QoS-related parameters have both subjective and objective elements in it [10]. The network attributes, for example, the importance of received signal strength and bandwidth are objective in nature. Application related attributes such as delay, packet loss, and jitter show some objectivity. However, some studies have already revealed their potential subjective natures. For example, a study conducted in Tanzania shows that the users give moderate importance to end-to-end delay over packet loss [11]. The study by ETSI reveals that the users give strong importance to end-to-end delay over packet loss [12]. Therefore, the importance of application-related performance parameters can vary based on changing contexts, for example, between home and industrial environments or urban and rural areas. The significance of applications can vary depending on the context as well. For example, an application related to the education services can have higher importance compared to one that provides some entertainment services. Moreover, the absence or presence of an application will affect the weights of others in the network.

For weight assignment, the available literature on QoS evaluation in network selection has mostly used the Analytic Hierarchy Process (AHP) method, which is primarily developed by Saaty [7, 8, 13]. Some studies have also assigned fixed weights to these parameters based on their importance to service performance [14]. Both AHP and fixed weight methods are unable to handle the subjective and ambiguous factors related to weight determination such as context-based significance. In this study, Fuzzy Analytical Hierarchy Process (FAHP) with the extent analysis method is applied to bring the context-based information into the picture. This method is capable of handling ambiguity in any particular subject. It is also possible to assign the weights dynamically to the relevant parameters by using this method.

## 3. THE UNIFIED QoS METRIC MEASUREMENT FUNCTIONS

The quality of service on any network or application is usually evaluated through a set of specific metrics. For example, to assess the performance of any voice application, the delay, jitter and packet loss are measured and compared with the acceptable values of these parameters. Similarly, the QoS of any network is evaluated through parameters such as delay, packet loss, throughput, and available bandwidth. The presence of different types of communication technologies and applications in a heterogeneous network makes its QoS assessment method a challenging task. To deal with such challenges, this paper introduces unified metric measurement functions.

The QoS evaluation approach proposed in this work considers any heterogeneous network as a set of three layers; these are the application layer, the radio access network layer and the network configuration layer. Each of these layers uses a function to quantify a unified QoS metric, which flows to the next layer and derive the combined metric of that layer. Figure 1 shows the flowchart of the proposed approach. In the application layer, a function is defined to derive the QoS of each application through a unified metric. This function combines the values of several application-related performance metrics. As such, the QoS of a network-based application is treated as a function of QoS-related parameters. This can be expressed as:

$$QoSAM_A = f\left(QP_1, QP_2, ..., QP_p\right) \qquad (1)$$

where $A$ denotes a network-based application, and $QP$ refers to the QoS-related parameters. Then in the radio access network layer or RAN layer, the QoS of each access network, which are present in the network, is evaluated. This evaluation is conducted based on the performances of the active applications in those access networks. Hence, the QoS of an access network is viewed as a function of the application QoS metrics. It can be expressed as:

$$QoSRM_R = f\left(QoSAM_{A_{i=(1,2,...,m)}}\right) \qquad (2)$$

where $R$ denotes any radio access network, and $i$ refers to the number of active applications present on a network. Finally, to evaluate the QoS of the overall network configuration another function is defined, which uses the radio access network metrics as its input. This can be expressed as:

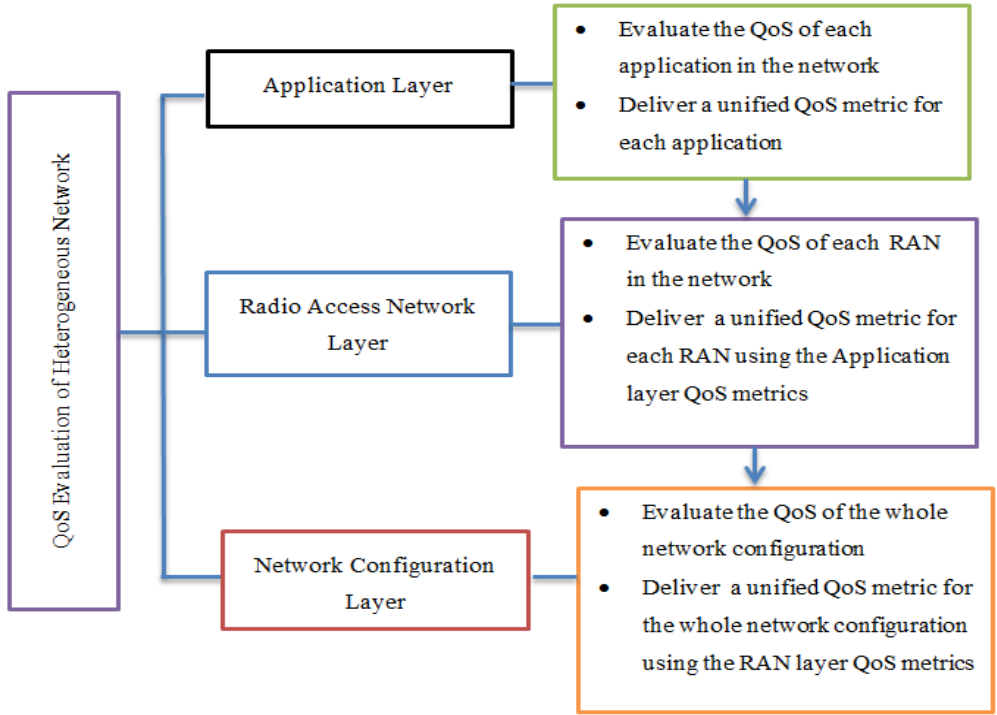

Figure 1. The concept of Unified QoS Metric

Table 1. Example of Application Weight Calculation

| Networks | Considered Parameters | | | Weights | |
|---|---|---|---|---|---|
| | Application | Service | Number of Users | $A_1$ | $A_2$ |
| $N_1$ | $A_1$ | Education | 20 | $w_{A_1}^{N_1}$ | $w_{B_1}^{N_1}$ |
| | $A_2$ | Entertainment | 18 | | |
| $N_2$ | $A_1$ | Health | 10 | $w_{A_1}^{N_2}$ | $w_{B_2}^{N_2}$ |
| | $A_2$ | Education | 6 | | |

$$QoSCM = f\left(QoSRM_{R_{j=(1,2,\ldots,n)}}\right) \quad (3)$$

where $j$ refers to the number of radio access networks present on a network.

## 4. The Application Weight Calculation Algorithm

The weights of applications are considered during the QoS evaluation of radio access networks. The radio access network metric can be used to label a network as a particular service-oriented network such as education or health by integrating the application weights. For instance, if the QoS metric of a radio access network that is mainly used for health services, is always good, then that network can be taken as a suitable health service-oriented one for future use.

The importance of the applications is subject to change depending on the requirements of particular networks. The service operators can update these criteria according to their particular circumstances. The criteria for this study have been formulated using the studies relevant to distance education-based service models [15]. For instance, in these service models, Videoconferencing (VC) bears more significance compared to voice-based applications. On the other hand, in a more general sense, VC may be less significant than voice-based applications as the latter is more easily amiable to the users. Therefore, the criteria to decide the applications weights in this regard can be the number of users using the application and the purpose and the context of that application usage.

Table 1 shows the example of two networks, $N_1$ and $N_2$, which have applications $A_1$ and $A_2$ with different number of users. The application weights are expressed as $w_{A_1}^{N_1}$, $w_{A_1}^{N_2}$, $w_{B_1}^{N_1}$, and $w_{B_2}^{N_2}$. In this paper, the weights of these applications are defined based on two criteria, the importance of the service, to which it belongs, and the number of users using that application. Other evaluation rules can be integrated based on individual needs. In the network $N_1$, it is considered that $w_{A_1}^{N_1} > w_{B_1}^{N_1}$. This is determined depending on the fact that in the network $N_1$, the application $A_1$ is used by more users than the application $A_2$ as the application $A_1$ is used for educational services, whereas, the application $A_2$ is used for entertainment services. Therefore, when the QoS metric in the network $N_1$ is calculated considering these application weights, the QoS metric value reflects the significance of the service the application provides. As a result, if the QoS value is good for the application $A_2$ and poor for the application $A_1$, the outcome of the QoS level of the network will be poor. That is because, the majority of the users in that network experience poor performance for a valuable service. If, in any case, the entertainment service application has more users, the result will also be same as the education service is set to have higher significance than the entertainment service. These findings can change based on specific network requirements.

On the other hand, the Network $N_2$ supports both education and health services. As more users are using the health services compared to the education services, the application weight of $A_1$ is greater than $A_2$, $w_{A_1}^{N_2} > w_{B_2}^{N_2}$. If the QoS value of the network $N_2$ is good, then it can be categorized as a health service-oriented network. Therefore, the configurations of $N_2$ can be recommended for any network that aims to deploy network-based health services in the future. Service operators can input these criteria to change the weights dynamically for any network.

The weight calculation involves two steps. At first, the alternatives, criteria, and the fuzzy judgement matrix are defined. Then in the second step, the actual weight is calculated based on those measures. FAHP-based calculations include: establishing a set of alternatives $X = \{x_1, x_2, \ldots, x_m\}$, a set of goal or evaluation criteria $G = \{g_1, g_2, \ldots, g_n\}$, a fuzzy judgement matrix (FJM), with elements $\widetilde{r_{ij}}$ that represents the relative importance of each pair of criteria $i$ and $j$, and a weighting vector $w = (w_1, w_2, \ldots, w_n)$. Both steps involve the concept of Triangular Fuzzy Number (TFN) and fuzzy addition and multiplication operations. To derive the FJM for the first step the importance scale presented in Table 2 is used. It shows the TFN $K_t = (l_t, m_t, u_t)$ where t=1, 2,…, 9, and $l_t, u_t$ and $m_t$ are the lower, upper and the middle value of the fuzzy number $K_t$ respectively. Table 3 shows the pair-wise comparison matrix for VC, voice, and VS applications formed based on the cited studies. The importance scale of Table 2 is used for the comparisons. If one of the applications is absent from the network, these pair-wise comparison matrices are subject to change.

For the second step of FAHP, different methods are proposed. The most prominent one is Chang's extent analysis method [16]. This method is chosen as it provides easy and flexible options for the weight calculation. The steps of the extent analysis method are as follows:

At first, the sums of the each row of the defined fuzzy comparison matrix are calculated. Then the normalization of the row sums is conducted using fuzzy multiplication to obtain fuzzy synthetic analysis. Therefore, in the fuzzy comparison matrix, the fuzzy synthetic analysis of criteria $G_i$ of alternative $X_m$ is calculated as:

$$D_{G_i}^{X_m} = \sum_{j=1}^{n} \tilde{r}_{ij} \otimes \left[ \sum_{i=1}^{n} \sum_{j=1}^{n} \tilde{r}_{ij} \right]^{-1}$$
$$= \left( \frac{\sum_{j=1}^{n} l_{ij}}{\sum_{i=1}^{n} \sum_{j=1}^{n} u_{ij}}, \frac{\sum_{j=1}^{n} m_{ij}}{\sum_{i=1}^{n} \sum_{j=1}^{n} m_{ij}}, \frac{\sum_{j=1}^{n} u_{ij}}{\sum_{i=1}^{n} \sum_{j=1}^{n} l_{ij}} \right) \quad (4)$$

where $i, j = \{1,2,3............n\}$ and $n$ is the number of criteria. In step 2, in order to rank the criteria against each alternative, the degree of possibility of two fuzzy numbers is applied. Therefore, $D_{G_2}^{X_m}(l_2, m_2, u_2) \geq D_{G_1}^{X_m}(l_1, m_1, u_1)$ is computed by the following equation:

Table 2. A FAHP-based Pair-wise Comparison Importance Scale

| Fuzzy Numbers | Definition | Triangular Fuzzy Number |
|---|---|---|
| $k_1(l_1, m_1, u_1)$ | Equal importance | (1,1,1) |
| $k_2(l_2, m_2, u_2)$ | Intermediate values | (1/2,3/4,1) |
| $k_3(l_3, m_3, u_3)$ | Moderate importance | (2/3,1,3/2) |
| $k_4(l_4, m_4, u_4)$ | Intermediate values | (1,3/2,2) |
| $k_5(l_5, m_5, u_5)$ | Strong importance | (3/2,2,5/2) |
| $k_6(l_6, m_6, u_6)$ | Intermediate values | (2,5/2,3) |
| $k_7(l_7, m_7, u_7)$ | Very strong importance | (5/2,3,7/2) |
| $k_8(l_8, m_8, u_8)$ | Intermediate values | (3,7/2,4) |
| $k_9(l_9, m_9, u_9)$ | Extreme importance | (7/2,4,9/2) |

$$V\left(D_{G_2}^{X_m} \geq D_{G_1}^{X_m}\right) = \sup \left[ \min \left( \mu_{D_{G_1}^{X_m}}(x), \mu_{D_{G_2}^{X_m}}(y) \right) \right] \quad (5)$$

It can be also expressed as:

$$V\left(D_{G_2}^{X_m} \geq D_{G_1}^{X_m}\right) = hgt\left(D_{G_2}^{X_m} \cap D_{G_1}^{X_m}\right)$$
$$= \mu_{D_{G_2}^{X_m}}(d) = \begin{cases} 1 & \text{if } m_1 \geq m_2 \\ 0 & \text{if } l_1 \geq l_2 \\ \frac{l_2 - u_2}{(m_2 - u_2) - (m_1 - l_1)} & \text{otherwise} \end{cases} \quad (6)$$

and

$$V\left(D_{G_1}^{X_m} \geq D_{G_2}^{X_m}\right) = hgt\left(D_{G_1}^{X_m} \cap D_{G_2}^{X_m}\right)$$

$$= \mu_{D_{G_1}^{X_m}}(d) = \begin{cases} 1 & \text{if} \quad m_1 \geq m_2 \\ 0 & \text{if} \quad l_2 \geq u_1 \\ \frac{l_2 - u_1}{(m_1 - u_1) - (m_2 - l_2)} & \text{otherwise} \end{cases} \quad (7)$$

Table 3. Pair-wise Comparison Matrix for different Applications

| Applications | Criteria | VC | | Voice | | VS | |
|---|---|---|---|---|---|---|---|
| VC | Purpose of Usage | (1, 1, 1) | | (3/2,2,5/2) | (1.09, 1.5, 2) | (2/3,1,3/2) | (0.84, 1.25, 0.75) |
| | Number of Users | (1, 1, 1) | | (2/3,1,3/2) | | (1,3/2,2) | |
| Voice | Purpose of Usage | (2/5,1/2,2/3) | (0.54, 0.75, 1.09) | (1, 1, 1) | | (2/3,1,3/2) | (1.59, 2, 2.5) |
| | Number of Users | (2/3,1,3/2) | | (1, 1, 1) | | (5/2,3,7/2) | |
| VS | Purpose of Usage | (2/3, 1,3/2) | (0.59, 0.84, 1.25) | (2/3,1,3/2) | (0.48, 0.67, 0.95) | (1, 1, 1) | |
| | Number of Users | (1/2,2/3,1) | | (2/7,1/3,2/5) | | (1, 1, 1) | |

where $d$ is the ordinate to validate if the highest intersection point $D$ is between $\mu_{D_{G_2}^{X_m}}$ and $\mu_{D_{G_1}^{X_m}}$. Both the values of $V\left(D_{G_2}^{X_m} \geq D_{G_1}^{X_m}\right)$ and $V\left(D_{G_1}^{X_m} \geq D_{G_2}^{X_m}\right)$ are required to compare $\mu_{D_{G_2}^{X_m}}$ and $\mu_{D_{G_1}^{X_m}}$. For large numbers of criteria, the degree of possibility is applied as:

$$V\left(D_{G_1}^{X_m} \geq D_{G_2}^{X_m}, D_{G_3}^{X_m}, \ldots, D_{G_n}^{X_m}\right) = V\left[\left(D_{G_1}^{X_m} \geq D_{G_2}^{X_m}\right) \text{ and } \left(D_{G_1}^{X_m} \geq D_{G_3}^{X_m}\right) \text{ and} \ldots \left(D_{G_1}^{X_m} \geq D_{G_n}^{X_m}\right)\right]$$

$$= \min V\left(d_{G_1}^{X_m} \geq d_{G_n}^{X_m}\right) \quad (8)$$

Assume that $d'\left(C_{G_n}^{X_m}\right) = \min V\left(d_{G_1}^{X_m} \geq d_{G_n}^{X_m}\right)$

In step 3, the weight vector **w** for each alternative is calculated. This is obtained as:

$$\mathbf{w}'_m = \left(d'\left(C_{G_1}^{X_m}\right), d'\left(C_{G_2}^{X_m}\right), \ldots, d'\left(C_{G_n}^{X_m}\right)\right)^T \quad (9)$$

In step 4, the normalized weight vector is calculated for each alternative as:

$$\mathbf{w}_m = \left(d\left(C_{G_1}^{X_m}\right), d\left(C_{G_2}^{X_m}\right), \ldots, d\left(C_{G_n}^{X_m}\right)\right)^T$$

$$= \left(\frac{d\left(C_{G_1}^{X_m}\right)}{\sum_{j=1}^{n} C_{G_n}^{X_m}}, \frac{d\left(C_{G_2}^{X_m}\right)}{\sum_{j=1}^{n} C_{G_n}^{X_m}}, \ldots, \frac{d\left(C_{G_n}^{X_m}\right)}{\sum_{j=1}^{n} C_{G_n}^{X_m}}\right) \quad (10)$$

The weight vector of the considered applications is calculated as:
$\mathbf{w}'_A(\text{VC}, \text{Voice}, \text{VS}) = (1, \ 0.94, \ 0.56)$

The normalization weight vector is as follows:
$\mathbf{w}_A = (w_{\text{VC}}, w_{\text{Voice}}, w_{\text{VS}}) = (0.4, 0.38, 0.224)$

# 5. SIMULATION STUDIES

This section presents some simulation studies and result analysis to investigate the impact of various factors on the performance of network-based applications which justify the concept of unified metrics. The factors, which are considered, are different environments, the presence of different traffic types in each RAN, and the number of active users in each RAN.

Figure 2 shows the average packet loss for 20 simultaneous voice calls in the UMTS network for different environments. In this case, the rural outdoor calls experience the highest amount of packet loss and the medium city outdoor calls experience the lowest amount of average packet loss. The small city outdoor calls experience 0.47% more packet loss than the medium city outdoor calls. In the mixed small city urban environment, when the calls take place between the outdoor and indoor office environment experience 5.68 % and 6.61% more packet loss than the urban indoor calls and the urban outdoor calls respectively. The rural outdoor calls experience 15.55% more packet loss than the small city outdoor calls.

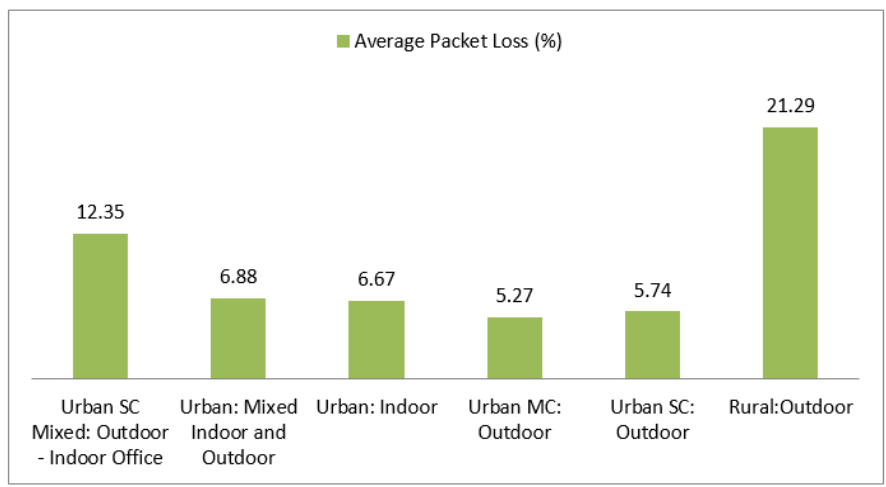

Figure 2. Average Packet Loss for 20 Voice calls in the UMTS Network

Figure 3 shows the average packet loss experienced by calls in different urban environments. The calls in the medium city (MC) outdoor environment undergo the lowest amount of packet loss. The pedestrian environment calls experience around 0.21% more packet loss than the calls in the urban indoor environment. The urban indoor calls experience around 1.4% and 0.93% more packet loss than the small and the medium city outdoor calls respectively.

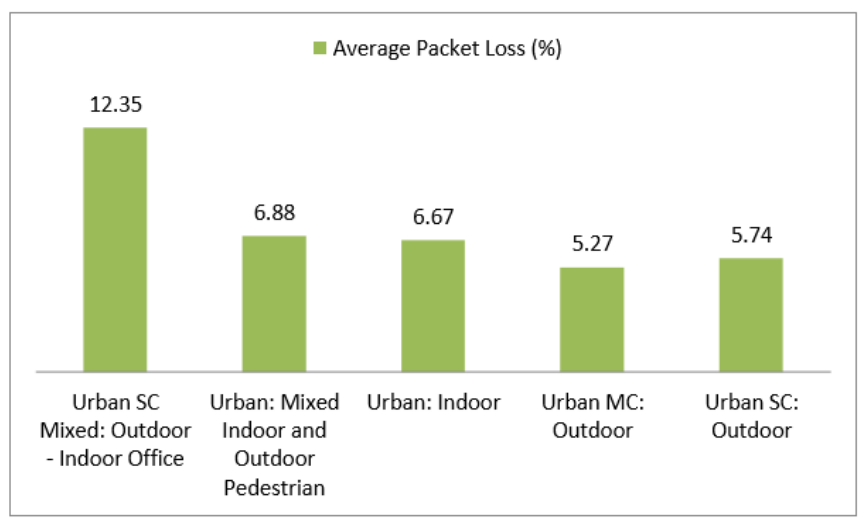

Figure 3. Average Packet loss for 20 Simultaneous Calls in Urban Environments

In the next stage of the simulations, the values of QoS-related parameters are analysed for the VS clients. The VS clients are placed on a UMTS network in the rural outdoor area, and the server is on a WiMAX network in an urban outdoor area. Figure 4 shows the packet loss that the clients experience in different environments. The second VS client in the rural outdoor environment experiences a 0.22% more packet loss than the VS client in the urban outdoor. The third VS client in the rural outdoor environment experiences a 0.31% packet loss than the client in the urban outdoor. These clients experience almost the same packet loss for the urban indoor and the urban outdoor environments.

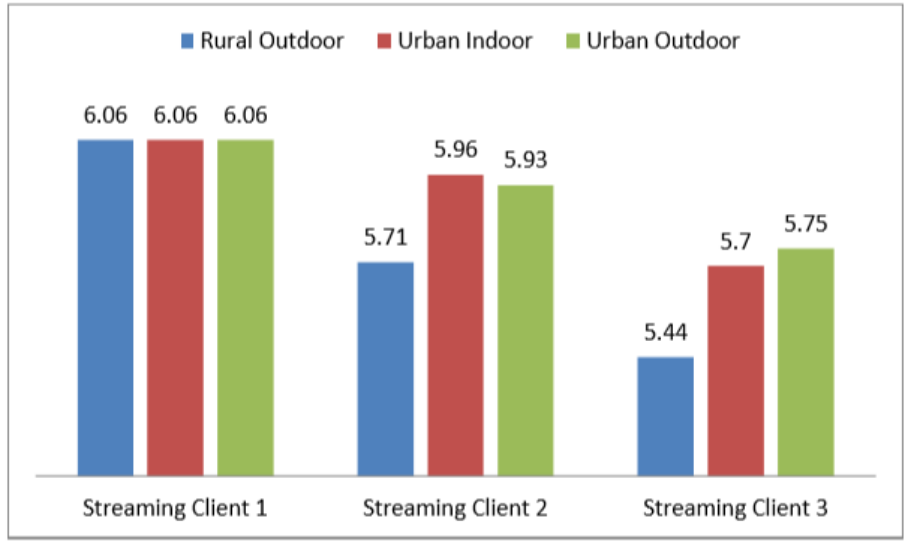

Figure 4. Packet loss for VS Clients in the UMTS-WiMAX Network

Then the impacts of technology, service, and the number of user-related factors on the application and network performance are analysed. Figure 5 shows the comparison of average packet loss for different number of voice calls. For eight new calls in the network, the callers in the rural outdoor environment experience 8.91% more average packet loss. In the small city outdoor environments, this loss is increased by 2.54%. Table 4 presents the percentage of increased packet loss for each environment.

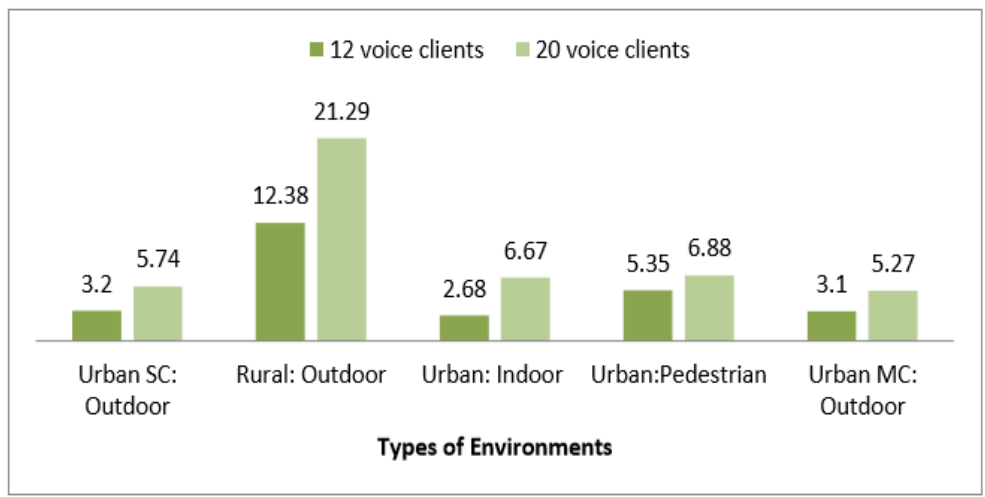

Figure 5. Impact of Number of Users on Average Packet loss in Different Environments

The above-discussed simulation result analysis show that in the wireless networks the application performance is affected by different factors. These factors could be environment, technology, network architecture, or traffic related. The results also illustrate that it is difficult to define any acceptable fixed value for these parameters due to involvement of multiple factors. Therefore, an acceptable range is more suitable which would consider all relevant factors and unified metric measurement oriented functions.

Table 4. QoS Parameters for VS Clients

| Path loss Model | | Average Packet Loss | Average End-to-end Delay (msec) | Average Delay Variation (msec) |
| --- | --- | --- | --- | --- |
| Client side | Server side | | | |
| Rural outdoor | Urban outdoor | 5.72 | 314.5 | 6.89 |
| Urban outdoor | Urban outdoor | 5.61 | 314.2 | 7.00 |
| Urban indoor | Rural outdoor | 6.15 | 314.5 | 6.89 |
| Rural outdoor | Suburban outdoor | 6.21 | 314.4 | 6.87 |

## 6. MEASURING THE Impact of Application Importance

This section evaluates the impacts of dynamic application weights on the unified QoS metrics. The performance of a few network-based service models have been analyzed using the fixed weight-based method in our previous work [17]. In this work, a detailed analysis is conducted to present the effects of dynamic weights on the performance of the same service models. The weights are calculated for each application according to the changing circumstances of the network. These weights are entered as inputs to derive the network QoS metric.

In the previous work, the voice application is set to have a higher importance than the VS application and the weights have been fixed as 0.6 and 0.4. In this work, those weights are set to change based on the pair-wise comparison matrices presented in Table 3. Figure 6 shows the QoS analysis of the scenario with twelve voice calls and one VS session on the network. The figure clearly indicates that when the voice and VS applications have equal importance, the access network has a good QoS level (e.g. 0.81). It shows an average QoS level (e.g. 0.62) for voice and a good QoS level (e.g. 1) for VS. When the importance level of voice application has been changed from having equal to extreme importance over VS application, the access network QoS comes down to an average value of 0.63. Although, the performance of the VS application is good, because of having a lower importance, it has less effect on the network QoS level. On the other hand, the voice application, being extremely important, has a greater impact on the network QoS level.

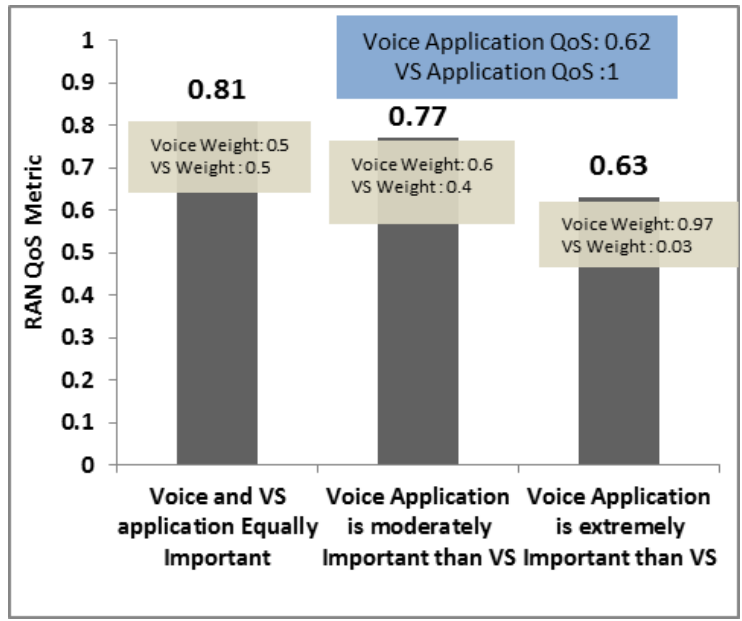

Figure 6. The Effect of Application Importance on Network QoS

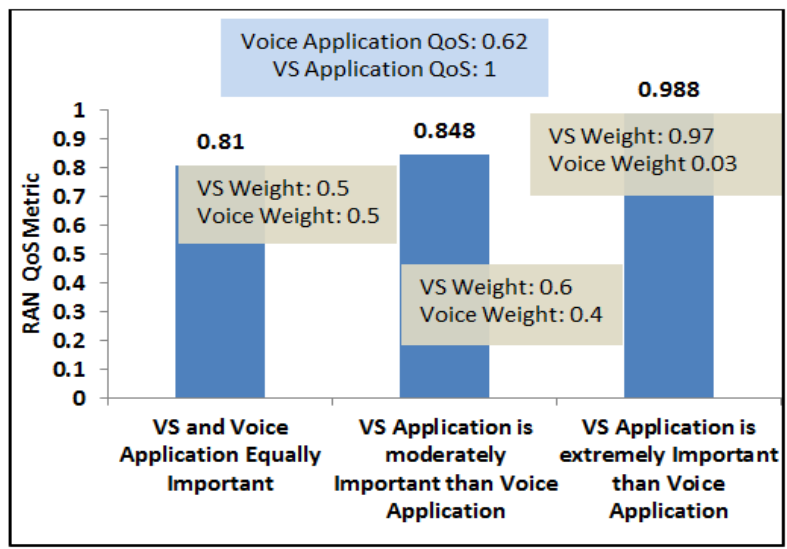

Figure 7. The Effect of Application Importance on the Network Performance

Figure 7 shows a similar type of analysis with the altered importance of voice and VS applications. When the VS application has extreme importance, the network QoS improves due to the impact of application weights on the network performance.

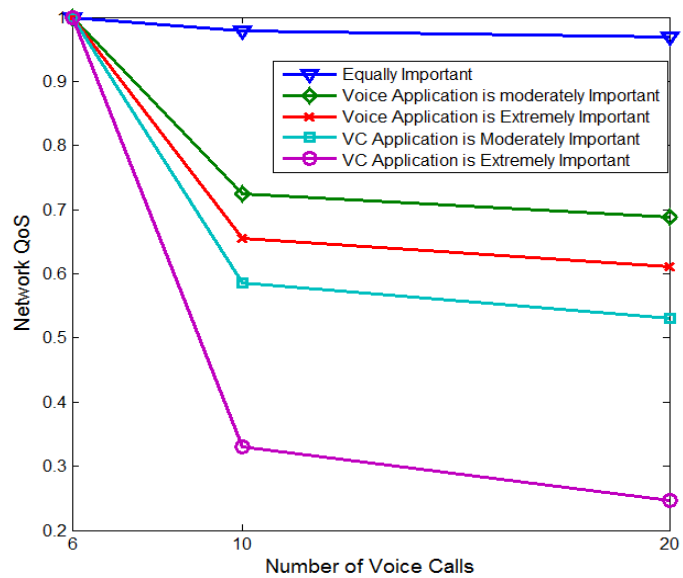

Figure 8. Network Performance Analysis with changing Application Importance

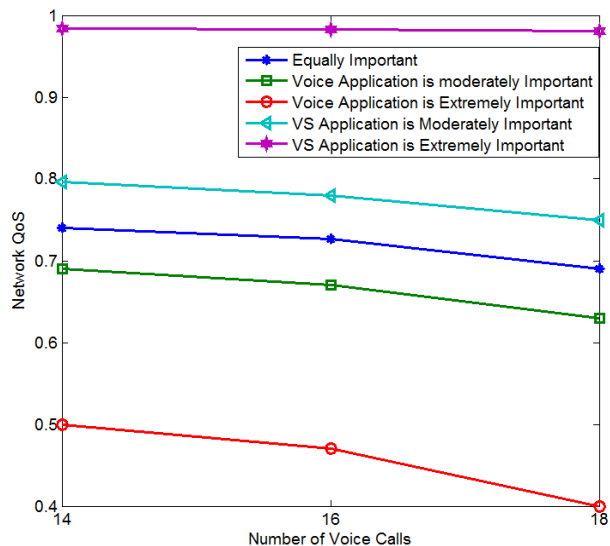

Figure 9. The Performance Analysis with different Application Significance

Figure 8 illustrates the QoS in a voice-based network for different number of calls when the importance of the application changes. When the VC application has extreme and moderate importance over voice application respectively, the network shows a poor QoS level. The reason is that the VC application with ten and twenty voice calls on the network experience a poor quality. On the other hand, the network takes an average QoS level with ten and twenty voice calls when the voice application has moderate and extreme importance over VC application respectively.

Figure 9 shows the QoS in the UMTS-WiMAX network for different number of calls when the significance of the application changes. When the VC application has extreme importance and moderate importance over voice application respectively, the network shows a poor QoS level. The reason is that the VC application with ten and twenty voice calls on the network experience a poor quality. On the other hand, the network takes an average QoS level with ten and twenty voice calls when the voice application has moderate and extreme importance over VC application respectively.

The above-discussed analyses clearly indicate that using application importance the QoS level can be adjusted to reflect the service-centric performance of any network. If an important application in a network experiences poor QoS, the overall network is recognized to have a poor performance regardless of the performance of other applications. In this way, the user experience or QoE can be integrated into the network QoS level as well. If the users experience poor performance for an important application, the overall network QoS reflects a lower QoS level. Therefore, it is easy to figure out the application for which the network is experiencing a poor performance.

## 7. CONCLUSIONS

In this paper, an application-based QoS analysis method has been proposed and evaluated. In assessing the overall QoS level of a heterogeneous network, the levels of importance of applications are included as weights. The key contributions of this work include the proposing of a methodical approach for calculating and applying these weights using the concept of unified metric measurement functions. Extensive simulation studies, utilizing these weights for QoS assessment of various heterogeneous configurations supporting a variety of applications, have also been carried out. These studies demonstrate how the inclusion of the application importance weights for QoS evaluations, can assist in a systemic choosing of a fitting network configuration. In our future works, we intend to include several other factors that can influence the QoS provisions of a heterogeneous network supporting real-time applications.

## REFERENCES

[1] Cisco, Cisco Visual Networking Index: Global Mobile Data Traffic Forecast Update, 2012–2017, Cisco, 2013.

[2] M. C. Lucas-Estañ and J. Gozalvez, "On the Real-Time Hardware Implementation Feasibility of Joint Radio Resource Management Policies for Heterogeneous Wireless Networks," *Mobile Computing, IEEE Transactions on*, vol. 12, no. 2, 2013, pp. 193-205; DOI 10.1109/TMC.2011.256.

[3] I. Lazar and M. Jude, "Network design and management for video and multimedia applications," 2008; http://searchenterprisewan.techtarget.com/feature/Network-design-and-management-for-video-and-multimedia-applications.

[4] M. Perlin, "Downtime, Outages and Failures - Understanding Their True Costs," 2013; http://www.evolven.com/blog/downtime-outages-and-failures-understanding-their-true-costs.html.

[5] S. M. Kantubukta Vasu, Sudipta Mahapatra, Cheruvu S Kumar, "QoS-aware fuzzy rule-based vertical handoff decision algorithm incorporating a new evaluation model for wireless heterogeneous networks," *EURASIP Journal on Wireless Communications and Networking,* vol. 2012, 2012.

[6] A. Alshamrani, S. Xuemin, and X. Liang-Liang, "QoS Provisioning for Heterogeneous Services in Cooperative Cognitive Radio Networks," *Selected Areas in Communications, IEEE Journal on*, vol. 29, pp. 819-830, 2011.

[7] A. Sgora, P. Chatzimisios, and D. Vergados, "Access Network Selection in a Heterogeneous Environment Using the AHP and Fuzzy TOPSIS Methods," in *Mobile Lightweight Wireless Systems*. vol. 45, P. Chatzimisios, C. Verikoukis, I. Santamaría, M. Laddomada, and O. Hoffmann, Eds., ed: Springer Berlin Heidelberg, 2010, pp. 88-98.

[8]  S. Qingyang and A. Jamalipour, "Network selection in an integrated wireless LAN and UMTS environment using mathematical modeling and computing techniques," *Wireless Communications, IEEE,* vol. 12, pp. 42-48, 2005.

[9]  W. Lusheng and D. Binet, "MADM-based network selection in heterogeneous wireless networks: A simulation study," in *Wireless Communication, Vehicular Technology, Information Theory and Aerospace & Electronic Systems Technology, 2009. Wireless VITAE 2009. 1st International Conference on*, 2009, pp. 559-564.

[10]  Z. Wenhui, "Handover decision using fuzzy MADM in heterogeneous networks," *Proc. Wireless Communications and Networking Conference, 2004. WCNC. 2004 IEEE*, 2004, pp. 653-658 Vol.652.

[11]  E. Sedoyeka, Z. Hunaiti, and D. Tairo, "Analysis of QoS Requirements in Developing Countries," *International Journal of Computing and ICT Research*, vol. 3, no. 1, 2009, pp. 18-31.

[12]  ETSI, Review of available material on QoS requirements of Multimedia Services, ETSI 2006.

[13]  E. Stevens-Navarro, L. Yuxia, and V. W. S. Wong, "An MDP-Based Vertical Handoff Decision Algorithm for Heterogeneous Wireless Networks," *Vehicular Technology, IEEE Transactions on*, vol. 57, no. 2, 2008, pp. 1243-1254; DOI 10.1109/tvt.2007.907072.

[14]  Wen-Tsuen and S. Yen-Yuan, "Active application oriented vertical handoff in next-generation wireless networks," *Proc. Wireless Communications and Networking Conference*, 2005 IEEE, 2005, pp. 1383-1388.

[15]  O. I. Hillestad, A. Perkis, V. Genc, S. Murphy, and J. Murphy, "Delivery of on-demand video services in rural areas via IEEE 802.16 broadband wireless access networks," *Proc. Proceedings of the 2nd ACM international workshop on Wireless multimedia networking and performance modeling*, ACM, 2006, pp. 43-52.

[16]  D.-Y. Chang, "Applications of the extent analysis method on fuzzy AHP," *European Journal of Operational Research*, vol. 95, no. 3, 1996, pp. 649-655; DOI http://dx.doi.org/10.1016/0377-2217(95)00300-2.

[17]  F. Farid, S. Shahrestani, and C. Ruan, "QoS analysis and evaluations: Improving cellular-based distance education," *Proc. Local Computer Networks Workshops (LCN Workshops), 2013 IEEE 38th Conference on*, 2013, pp. 17-23.

**Authors**

Farnaz Farid is pursuing her PhD degree in Information Technology and Communications at the Western Sydney University.  Prior to that she has worked in China as a web application developer and web business SME at IBM. Her research interests include wireless and cellular networking, web engineering, and technology for development**.**

Seyed Shahrestani completed his PhD degree in Electrical and Information Engineering at the University of Sydney. He joined  Western Sydney University in 1999, where he is currently a Senior Lecturer. He is also the head of the Networking, Security and Cloud Research (NSCR) group at Western Sydney University.

Chun Ruan received her PhD degree in Computer Science in 2003 from the University of Western Sydney. Currently she is a lecturer in the School of Computing, Engineering and Mathematics at Western Sydney University. Prior to that, she worked as an associate professor, lecturer and associate lecturer at the Department of Computer Science, Wuhan University, China.